\documentclass[preprint,aps,prd,12pt,preprintnumbers,nofootinbib,superscriptaddress]{revtex4-2}
\usepackage{xcolor}
\usepackage{braket}
\usepackage{amssymb,amsmath,amsfonts,comment,latexsym,graphicx,epstopdf,float}
\usepackage{color}
\usepackage{fancyvrb}
\usepackage{chngcntr}
\usepackage{etoolbox}
\usepackage{ulem}
\usepackage{array}
\usepackage{bbold}
\usepackage{subfigure}
\usepackage{slashed}
\usepackage{multirow}

\usepackage[colorlinks=true,citecolor=blue,linkcolor=red,filecolor=purple,urlcolor=magenta]{hyperref}

\usepackage{float}

\newcommand{\be}{\begin{equation}}
\newcommand{\ee}{\end{equation}}
\newcommand{\ba}{\begin{eqnarray}}
\newcommand{\ea}{\end{eqnarray}}

\newcommand{\grts}{\raise.3ex\hbox{$>$\kern-.75em\lower1ex\hbox{$\sim$}}}
\newcommand{\lets}{\raise.3ex\hbox{$<$\kern-.75em\lower1ex\hbox{$\sim$}}}

\newcommand{\dd}{\text{d}}

\usepackage{color}
\usepackage{amssymb,amsmath,amsfonts}
\usepackage{epstopdf} 
\usepackage{braket}

\usepackage{autobreak}

\begin{document}
%
%
\title{\vspace*{0.5in} 
New models of nonsingular black hole dark matter \\ from limiting curvature
\vskip 0.1in}
\author{Selin A\c{s}mano\u{g}lu}\email[]{sasmanoglu@wm.edu}
\affiliation{High Energy Theory Group, Department of Physics, William \& Mary, Williamsburg, VA 23187-8795, USA} 
\author{Jens Boos}\email[]{jens.boos@kit.edu}
\affiliation{Institute for Theoretical Physics, Karlsruhe Institute of Technology, D-76128 Karlsruhe, Germany}
\author{Christopher D. Carone}\email[]{cdcaro@wm.edu}
\affiliation{High Energy Theory Group, Department of Physics, William \& Mary, Williamsburg, VA 23187-8795, USA}
\date{September 15, 2025}
%
%

\begin{abstract}
We consider phenomenological models for nonsingular black holes that satisfy the limiting curvature condition (i.e., that have curvatures that are always sub-Planckian in size) while having a more general dependence on the black hole mass than the most studied examples. These models allow black holes to exist while having regulators that are larger than the horizon scale;  it has been shown previously that this can lead to observable consequences in an astrophysical setting, for allowed choices for the regulator scale. Noting that substantial horizon-scale modifications of the metric will affect black hole thermodynamics and Hawking radiation, we study these metrics in the context of primordial black hole dark matter. Considering examples with de\,Sitter and Minkowski cores, respectively, we study the effect of the regulator in these metrics on the allowed black hole mass ranges (or ``bands"), the black hole temperature, specific heat and lifetime, and the bounds on the primordial black hole fraction of the total dark matter density from the observed extragalactic gamma ray background.
\end{abstract}

\maketitle
\newpage
\section{Introduction} \label{sec:intro}

Particle dark matter has been a leading contender for explaining the observed galactic rotation curves~\cite{Zwicky:1933gu,Rubin:1970zza}, the origin of structure formation in the universe~\cite{Frenk}, the power spectrum of the cosmic microwave background radiation~\cite{WMAP:2010qai}, and gravitational lensing around astrophysical objects like the Bullet Cluster~\cite{Clowe:2006eq}. Two elegant possibilities, weakly interacting massive particles (WIMPS)~\cite{Lee:1977ua} and axions~ \cite{Preskill1983,Abbott1983,Dine1983} may, in principle, be directly detected via a range of experimental means. For example, in the case of WIMPS, this includes searches for the nuclear recoil resulting from dark-matter-nucleon elastic scattering, direct production of dark matter via collider processes, and searches for byproducts of dark matter annihilation that occurs in space~\cite{Bozorgnia:2024pwk}. So far, all experimental searches for dark matter have returned null results, leading only to upper limits on the strength of the dark matter coupling to ordinary matter as a function of its mass~\cite{ParticleDataGroup:2024cfk}. While particle dark matter remains a compelling possibility, the absence of experimental evidence motivates consideration of other possibilities.

Primordial black hole dark matter is one such option that has gained some attention in the recent literature~\cite{Khlopov:2008qy,Carr:2016drx,Carr:2020gox,Carr:2020xqk,Carr:2025auw}. Black holes are predicted by general relativity and are known to exist from astrophysical observations~\cite{Bolton1972,Webster1972,Genzel:1966zz,Ghez:2000ay,LIGOScientific:2016aoc,EventHorizonTelescope:2019dse}, including the gravitational wave signals of black hole mergers~\cite{LIGOScientific:2016aoc} and the direct imaging of black hole shadows~\cite{EventHorizonTelescope:2019dse}.  It is not a radical extrapolation to assume that black holes may occur with masses much less than the solar mass scale, and there are known mechanisms for the production of primordial black holes in the early universe~\cite{Carr:1974nx,Khlopov:1985fch,Ivanov:1994pa}.  Black hole dark matter does not require the existence of an (unobserved) particle dark sector, or (unobserved) mediator particles that are needed in models that rely on the freeze-out mechanism to obtain the present dark matter relic density. In that sense, it is a scenario that is quite economical.  In this paper, we consider a specific class of black holes as possible candidates for primordial black hole dark matter.

One of the features of black holes that is problematic is the existence of a physical singularity in the black hole interior where curvature invariants become infinitely large and tidal forces diverge.  This is believed to be an indication of the breakdown of the classical description of gravity, and it is usually assumed that a theory of quantum gravity will resolve the singularity (i.e. render the curvature finite and smooth at that point).  However, there is no universally excepted quantum theory of gravity, nor is there a robust prediction of exactly how the divergence in the metric should be physically regulated.  Hence,  phenomenological metrics aimed at capturing some of the qualitative features of the ultraviolet physics have been used to study black holes that are free of singularities; these have been called ``regular black holes'' or ``nonsingular black holes'' in the literature \cite{Frolov:2016pav,Carballo-Rubio:2025fnc}; we adopt the latter term in the present work.

Generally, phenomenological metrics do not correspond to solutions of the  Einstein equations for simple energy-momentum tensors. While exact non-vacuum solutions have been obtained in the context of nonlinear electrodynamics \cite{Ayon-Beato:1998hmi,Ayon-Beato:1999qin,Ayon-Beato:1999kuh}, their Lagrangian does not reduce smoothly to the Maxwellian one, and can hence lead to theoretical inconsistencies \cite{Bronnikov:2000yz,Bronnikov:2000vy}. More recent developments include exact vacuum solutions in a class of higher-dimensional gravity theories called quasi-topological \cite{Bueno:2024eig,Bueno:2024dgm,DiFilippo:2024mwm}, which, in a symmetry-reduced setting, may also be applied directly to four spacetime dimensions \cite{Frausto:2024egp,Frolov:2024hhe}. These shortcomings are not necessarily a problem, since nonsingular black hole metrics are intended to reflect physics that goes beyond the classical description of gravity; the naive expectations of that classical theory, including energy conditions on the energy-momentum tensor, may not apply. Moreover, in the presence of modified field equations in a more fundamental theory of gravity, the precise form of the singularity theorems may likewise be altered. For these reason, these considerations are generally set aside in phenomenological studies.

The observational implications of nonsingular black holes with astrophysical masses have been considered in the literature~\cite{Bambi2023,Brahma:2020eos,Li:2024ctu,Liu:2020ddo,Chen:2019iuo}, as well as the possibility that nonsingular black holes may be primordial black hole dark matter~\cite{Calza:2024fzo,Davies:2024ysj,Calza:2024xdh,Calza:2025mwn} or drivers of cosmic acceleration~\cite{Dialektopoulos:2025mfz}. Some nonsingular black holes have been developed as an alternative to those with a minimum spherical radius~\cite{Bronnikov:2024izh,Bolokhov:2024sdy}, and their quasinormal ringing has also been studied~\cite{Skvortsova:2024wly}.  Since the masses of primordial black holes can be much smaller than the masses of astrophysical black holes, and the curvatures near the horizon are correspondingly larger, one might expect that modifications of general relativity will have more important effects in the vicinity of such primordial black holes.

In the present work, we focus on nonsingular black holes that have not been considered previously as candidates for primordial black hole dark matter and that are more general than the most-studied examples. The metrics we consider are guided by the limiting curvature condition~\cite{Markov:1982,Markov:1984ii,Polchinski:1989ae,Mukhanov:1991zn}, which elevates the conjecture that curvature invariants may not exceed maximum values set by the Planck scale to a universal principle.\footnote{Limiting curvature has been implemented at the Lagrangian level in general relativity \cite{Frolov:2021vbg} and nonsingular black hole solutions have been found in $(1+1)$ dimensions \cite{Trodden:1993dm,Easson:2017pfe,Frolov:2021kcv} and $(3+1)$ dimensions \cite{Frolov:2021afd}.} For concreteness, consider the well-known Hayward metric~\cite{Hayward:2005gi}
\begin{equation}
\dd s^2 = - F(r) \,\text{d}t^2 + \frac{1}{F(r)}\, \dd r^2 + r^2\, \dd\Omega^2\,\, , \hspace{3em} F(r)=1-\frac{2\, G M}{r} \frac{r^3}{r^3+L^3} \,\, ,
\label{eq:metric}
\end{equation}
where $L^3 \equiv 2\, G M  \ell^2$ and $\ell$ is a free parameter.  (Here we work in units where $c=\hbar=1$.)  For $\ell \neq 0$, there is no singularity and, for example, the Ricci scalar at the origin is ${\cal O} (1/\ell^2)$.  Provided that $\ell$ is greater than the Planck length $\ell_{P}$, the limiting curvature condition is satisfied.  It was noted in Ref.~\cite{Boos:2023icv} that a more general form of the mass dependence in the quantity $L^3$ is consistent with the limiting curvature condition and can lead to interesting phenomenological consequences.  In particular, kilometer-scale values of the regulator were shown to be phenomenologically viable, leading to horizon-scale modifications of astrophysical black holes, and new allowed black hole mass ranges ({\it i.e.}, where black holes exist rather than compact, horizonless objects).  These mass ranges were called ``black hole bands" in Ref.~\cite{Boos:2023icv}.  The following modification of $L^3$ was considered in that work,
\begin{equation}
L^3 = 2\, G M \ell^2 f(\hat{\ell}), \,\,\,\, \mbox{ with } \,\,\,\, \hat{\ell} \equiv \frac{\ell}{2\, G M} \,\, ,
\label{eq:ourL}
\end{equation}
where the chosen form of the dimensionless function $f(\hat{\ell})$ defines a given model.  This leads to the following finite curvature invariants for  $r \ll \ell$,
\begin{align}
R &= g{}^{\mu\nu}R{}_{\mu\nu} = \frac{12}{f(\hat{\ell}) \, \ell^2 } + {\cal O}(r^3)  \, , \\
 C^2 & = C{}_{\mu\nu\rho\sigma}C{}^{\mu\nu\rho\sigma} = \frac{12}{(GM)^2} \frac{r^6}{f(\hat{\ell})^4\,\ell^8} + {\cal O}(r^9)\,\, ,  \label{eq:weyl} \\
S^2 &= R_{\mu\nu} R^{\mu\nu} - R^2/4 = \frac{81}{4\, (G M)^2} \frac{r^6}{f(\hat{\ell})^4\,\ell^8}+{\cal O}(r^9) \,\, , \label{eq:ssq}
\end{align}
where $R$ is the Ricci scalar, $C{}_{\mu\nu\rho\sigma}$ denotes the Weyl tensor, and the invariant $S^2$ is the square of the tracefree Ricci tensor $S{}_{\mu\nu} = R{}_{\mu\nu}-(1/4)R \,g{}_{\mu\nu}$. The new black hole metrics that are defined via the introduction of the function $f(\hat{\ell})$ are subject only to the requirement that $\ell f(\hat{\ell})^{1/2}$ remains larger that the Planck length for any value of the mass $M$.
For example, the choice
\begin{equation}
f(\hat{\ell}) = \frac{1}{1+\hat{\ell}^4} + \epsilon
\label{eq:fdef}
\end{equation}
has the desired properties.  Imagining for the moment that $\epsilon \ll 1$,  $f(\hat{\ell}) \rightarrow 1+\epsilon \approx 1$ as $M \rightarrow \infty$, while it uniformly decreases to $\epsilon$ as $M\rightarrow 0$.   As long as $\sqrt{\epsilon}\, \ell > \ell_{P}$, the limiting curvature condition is satisfied for any $M$.\footnote{The $\epsilon$ factor was omitted from the discussion of  Ref.~\cite{Boos:2023icv};  $\epsilon$ may be taken small enough to have a negligible effect on the astrophysical observables considered in Ref.~\cite{Boos:2023icv} while still satisfying the limiting curvature condition.}  To see this, we note that despite the explicit $M$ dependence in Eqs.~(\ref{eq:weyl}) and (\ref{eq:ssq}), the maximum values of the curvature invariants are functions of $f(\hat{\ell})\, \ell^2$ only: 
\begin{align}
R_\text{max} &= \frac{12}{f(\hat{\ell})\, \ell^2} \, , \\
C^2_\text{max} &  = \frac{48(53-20\sqrt{7})}{(4-\sqrt{7})^6} \frac{1}{f(\hat{\ell})^2\, \ell^4} \approx \frac{0.66}{f(\hat{\ell})^2\, \ell^4} \, , \\
S^2_\text{max} &= \frac{16}{9 f(\hat{\ell})^2\ell^4} \approx  \frac{1.78}{f(\hat{\ell})^2\, \ell^4} \, . 
\end{align} 
The maxima are characterized by the length scale $f(\hat{\ell})^{1/2}\, \ell$, up to constants of $\mathcal{O}(1)$, and have only a very mild $M$-dependence via $f(\hat{\ell})$.

The introduction of the function $f(\hat{\ell})$ generalizes a mass dependence that is already present in the quantity $L^3$ that appears in the Hayward metric, but is a less conventional choice since it also introduces a mass dependence in the curvature invariants evaluated at $r=0$. In the absence of a rigorous derivation of regular black hole metrics from an underlying theory, it seems reasonable to allow for this possibility, provided that the modifications are consistent with the limiting curvature condition for all $M$.   We will also consider a similar modification of a nonsingular black hole metric that has vanishing curvature at the origin, one with phenomenological consequences that are similar to the nonsingular black holes that were discussed in  Ref.~\cite{Boos:2023icv}; in the example we study, the curvature invariants away from the origin satisfy the limiting curvature condition for all $M$. One of the interesting features of the allowed black hole mass bands in the models we consider is that the black holes will Hawking radiate only until their mass hits the edge of a band where its temperature vanishes and the black hole becomes a compact, horizonless object.  This will have implications for primordial black hole dark matter as we will discuss later in the present work.

Our paper is organized as follows.   In Sec.~\ref{sec:models}, we define the nonsingular black hole metrics of interest, and briefly review some of the phenomenological features that were pointed on in Ref.~\cite{Boos:2023icv}.  We will include examples where the curvature at the origin is finite with a nontrivial mass dependence, and where it is exactly vanishing.  In Sec.~\ref{sec:thermo} we discuss some of the thermodynamic features of these black holes and provide some simple estimates for their lifetimes.  In Sec.~\ref{sec:gamma}, we consider the bounds from the observed extragalactic gamma ray background on the fraction of the total dark matter density that may be primordial black holes of the type of interest.  In Sec.~\ref{sec:conc} we summarize our conclusions.

\section{Nonsingular Black Hole Models} \label{sec:models}

In this section, we define the nonsingular black holes of interest, which have either de\,Sitter or Minkowski cores. Nonsingular de\,Sitter-core black holes are a generic consequence of the quantum field theory corrections in curved spacetime to the Schwarzschild metric~\cite{Poisson:1988wc,Balbinot:1990zz}. Mechanisms that generate a positive effective vacuum energy at high curvature arise naturally in many semiclassical and quantum-gravity-motivated models and include quantum phase transitions~\cite{Dymnikova:2004zc,Mazur:2004fk}, condensate formation~\cite{Chapline:2000en}, or loop-quantum-gravity-induced corrections~\cite{Modesto:2004xx,Rovelli:2014cta}. While nonsingular black holes with anti-de\,Sitter (AdS) cores represent another logical possibility, they have been less frequently studied. AdS cores require sustained negative energy densities and finely tuned matching layers to remain static, which typically demands exotic matter sources not supported by known physics and are therefore considered less plausible~\cite{Ansoldi:2008jw,Lemos:2008cv,Bambi:2013caa}.  Hence, we do not consider that possibility in the present work.   However, we do consider a nonsingular black hole with a Minkowski core, as an example where the limiting curvature condition is satisfied everywhere with vanishing curvature invariants at the origin.  One should keep in mind that realistic physical realizations of Minkowski-core black holes may require fine tuning that may not be needed in models where the energy density near the origin is positive.

\subsection{de\,Sitter cores} \label{sec:desitter}
\begin{figure}[!htb]
\centering
\includegraphics[width=0.75\textwidth]{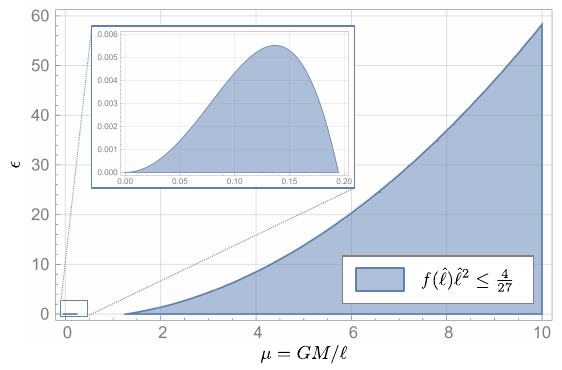}
\caption{Horizon structure in the de\,Sitter core example. The shaded regions correspond to parameter choices
for which horizons exist.}
\label{fig:horizons-hayward}
\end{figure}

The nonsingular black holes that were proposed in Ref.~\cite{Boos:2023icv} may be described as having de\,Sitter cores.
The metric function $F(r)$ may be expanded about the origin
\begin{equation}
F(r) = 1 - \frac{\Lambda \, r^2}{3} + {\cal O}(r^5) \,\, ,
\end{equation}
where $\Lambda \equiv  6 \, G M/L^3$ and where $L^3$ is defined by Eqs.~(\ref{eq:ourL}) and (\ref{eq:fdef}) for the black holes of interest to us.   This approximates de\,Sitter space with $\Lambda$ identified as the positive cosmological constant.   As noted in Ref.~\cite{Boos:2023icv}, horizons exist provided that
\begin{equation}
\hat{\ell}^2 f(\hat{\ell}) \leq \frac{4}{27}\,\, ,
\end{equation}
which, given Eq.~(\ref{eq:fdef}), has two solutions when $\epsilon=0$,
\begin{equation}
G M \alt 0.20 \,\ell  \,\,\,\,\,\mbox{ and }\,\,\,\,\, G M \agt 1.28\, \ell \,\, ,
\label{eq:branches}
\end{equation}
or
\begin{equation}
0.04 \, \ell \alt G M \alt 0.19 \,\ell  \,\,\,\,\,\mbox{ and }\,\,\,\,\,  G M \agt 1.29\, \ell \,\, ,
\label{eq:branches2}
\end{equation}
when $\epsilon=0.001$, a value we assume for definiteness in our later numerical analysis. The lower mass range is not present in the Hayward metric, {\it i.e.} in the case where $f(\hat{\ell})=1$, so that there are substantially relaxed lower bounds on black hole masses for the de\,Sitter-core metrics of interest to us, depending on how small one wishes to take $\epsilon$ consistent with the limiting curvature condition. See Fig.~\ref{fig:horizons-hayward} for a visualization of the black hole band structure for this type of nonsingular de\,Sitter core black hole.

 One of the main points of Ref.~\cite{Boos:2023icv} was that deviations from general relativity can be negligible for masses relevant in precise terrestrial experiments, but may nonetheless be significant for astrophysical black holes, due to the nontrivial mass dependence in $f(\hat{\ell})$.   This allows the regulator $\ell$ to be, for example, kilometer-scale and comparable to the horizon scale for astrophysical black holes, leading to observable consequences.  For example, the effect on the black hole shadow was found to be up ${\cal O}(5\%)$ relative to the expectation for a Schwarzschild black hole~\cite{Boos:2023icv}.  

In the present work, we will be concerned with black holes living in the new, lower-mass branch of Eq.~(\ref{eq:branches2}).  For primordial black hole dark matter, the regulator values that are comparable to the horizon scale are femtometer in size, so laboratory bounds on deviations from Newtonian gravity will not be relevant.  Nevertheless, the form of the metric and dependence on the mass and regulator will impact the dark matter phenomenology, as we will see  in Secs~\ref{sec:thermo} and \ref{sec:gamma}.

\subsection{Minkowski cores}

\begin{figure}[!htb]
\centering
\includegraphics[width=0.75\textwidth]{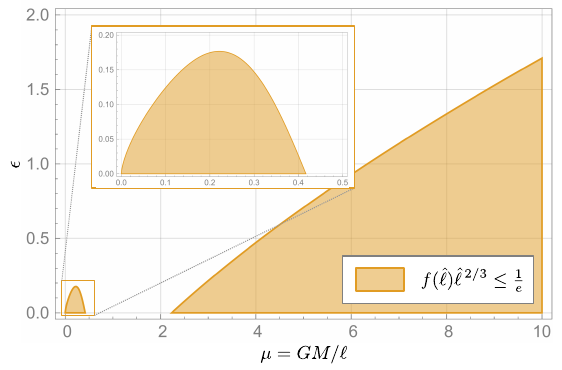}
\caption{Horizon structure in the Minkowski core example. The shaded regions correspond to parameter choices
for which horizons exist.}
\label{fig:horizons-minkowski}
\end{figure}

Another possibility lies in identically vanishing curvature at the origin, otherwise subjected to the limiting curvature condition. We consider the choice
\begin{equation}
F(r)=1-\frac{2 G M}{r} \exp\left[-\frac{(2GM \ell^2)^{1/3}}{r}\, f(\hat{\ell}) \right] \, ,
\label{eq:MCF}
\end{equation}
where $f(\hat{\ell})$ is again given by Eq.~\eqref{eq:fdef}. Computing the curvature invariants $R$, $C^2$, and $S^2$ close to the origin $r=0$ we find
\begin{align}
R &= \exp\left[ -\frac{(2GM\ell^2)^{1/3}}{r} \, f(\hat{\ell}) \right] \frac{(2GM)^{5/3} f(\hat{\ell})^2 \ell^{4/3} }{r^{5}} \qquad \text{(exact)} \, , \\
C^2 &= \exp\left[ -2\frac{(2GM\ell^2)^{1/3}}{r} \, f(\hat{\ell}) \right] \left[ \frac{(2GM)^{10/3} f(\hat{\ell})^4 \ell^{8/3} }{3 \, r^{10}} + \mathcal{O}(r^{-9}) \right] \, , \\
S^2 &= \exp\left[ -2\frac{(2GM\ell^2)^{1/3}}{r} \, f(\hat{\ell}) \right] \left[ \frac{(2GM)^{10/3} f(\hat{\ell})^4 \ell^{8/3} }{4 \,r^{10}} + \mathcal{O}(r^{-9}) \right] \, ,
\end{align}
which all vanish as $r \rightarrow 0$, justifying the notion of a ``Minkowski core.'' The maximum values are assumed for intermediate values of $r$, and for our parametrization take the explicit form
\begin{align}
R_\text{max} &= \frac{3125}{e^5} \frac{1}{f(\hat{\ell})^3 \ell^2} \approx \frac{21.06}{f(\hat{\ell})^3\ell^2} \, , \\
C^2_\text{max} &= \frac{64(216401 + 68432\sqrt{10})e^{-2(4+\sqrt{10})}}{3} \frac{1}{f(\hat{\ell})^6 \ell^4} \approx \frac{5.55}{f(\hat{\ell})^6\ell^4} \, , \\
S^2_\text{max} &= \frac{8796093022208 e^{-9-\sqrt{17}}}{(9-\sqrt{17})^9} \frac{1}{f(\hat{\ell})^6 \ell^4} \approx \frac{11.26}{f(\hat{\ell})^6\ell^4} \, .
\end{align}
These again are only mildly $M$-dependent via the function $f(\hat{\ell})$. Similar to the case of the de\,Sitter core, a black hole horizon only exists provided
\begin{align}
\hat{\ell}^{2/3} f(\hat{\ell}) \leq \frac{1}{e} \approx 0.3679 \, .
\end{align}
For $\epsilon=0$ this implies
\begin{align}
\label{eq:MCbranches-epsilon-zero}
GM \lesssim 0.4169 \, \ell \, \quad \text{and} \quad GM \gtrsim 2.2324 \, \ell \, ,
\end{align}
whereas for $\epsilon=0.001$ one finds
\begin{align}
0.00007086 \, \ell \lesssim GM \lesssim 0.4163 \, \ell \, \quad \text{and} \quad GM \gtrsim 2.2358 \, \ell \, ,
\label{eq:MCbranches-epsilon-nonzero}
\end{align}
where again the lower band emerges for $f(\hat{\ell}) \not= 1$.   Note that if one were to replace the quantity $(2 G M \ell^2)^{1/3} f(\hat{\ell})$ in Eq.~(\ref{eq:MCF}) by a mass-independent parameter with units of length, one would obtain the nonsingular black hole metric that has been studied previously in Refs.~\cite{Culetu:2013fsa,Culetu:2014lca,Ghosh:2014pba,Simpson:2019mud}; the mass dependence that we have introduced in Eq.~(\ref{eq:MCF}) not only assures that the limiting curvature condition is valid for all $M$, but also leads to the possibility of large regulator values and new black hole mass ranges, similar to the de\,Sitter core model defined by Eqs.~(\ref{eq:metric}), (\ref{eq:ourL}) and (\ref{eq:fdef}). See Fig.~\ref{fig:horizons-minkowski} for a visualization of this band structure.  As in our earlier de\,Sitter core model, the femtometer-sized regulators that are relevant for primordial black hole dark matter do not lead to observable deviations from Newtonian gravity in the laboratory or at larger length scales.

\section{Thermodynamics and lifetimes} \label{sec:thermo}

If a dark matter candidate is unstable, its decay rate must be computed to determine whether there will be a sufficient dark matter relic abundance in the universe at the present time.   For primordial black hole dark matter, this requires a consideration of the thermodynamic properties of the black holes of interest, including their thermal stability; black hole thermodynamics was not discussed in our earlier work on nonsingular astrophysical black holes, where it is less relevant~\cite{Boos:2023icv}.  In this section, we consider the thermodynamic properties of the de\,Sitter and Minkowski core black holes that were defined in the previous sections and we and provide estimates of their lifetimes as a function of the black hole mass.

In the absence of a concrete ultraviolet theory giving rise to such metrics, in what follows we shall assume that their relevant thermodynamic properties can be computed in analogy to general relativity.\footnote{See Ref.~\cite{Boos:2023pyr}, and references therein, for additional considerations regarding the thermodynamic properties of nonsingular black holes.} Namely, we assume that for static, spherically symmetric and asymptotically flat black hole metrics the Hawking temperature is given by the standard expression
\begin{equation}
\label{eq:temperature}
T_H \equiv \left. \frac{F'(r)}{4\pi} \right|_{r = r_H} \, ,
\end{equation}
where $F(r_H) = 0$ denotes the location of the outer horizon. In the present case we find for the class of de\,Sitter core nonsingular black holes
\begin{align}
\label{eq:rh-hayward}
r_H = \frac{2GM}{3}\left\{ 1 + 2\cos\left[ \frac13 \arccos \left( 1 - \frac{27}{2} \hat{\ell}^2 f(\hat{\ell}) \right) \right] \right\} \quad \text{(de\,Sitter core )} \, ,
\end{align}
whereas for the Minkowski core nonsingular black hole we calculate
\begin{align}
\label{eq:rh-minkowski}
r_H = 2GM \, \exp \left\{ W\left[ -\hat{\ell}^{2/3} f(\hat{\ell}) \right] \right\} \, , \quad \text{(Minkowski core)} \, ,
\end{align}
where $W(z)$ denotes the Lambert W function defined via $W(z) \exp[W(z)] = z$. After computing the Hawking temperature via \eqref{eq:temperature} the heat capacity can then be determined,
\begin{align}
C = \frac{\partial M}{\partial T_H} = \frac{1}{\frac{\partial T_H}{\partial M}} \, ,
\end{align}
which can be computed taking into account the $M$-dependence that appears explicitly in $F'(r)$, as well as in $r_H$ via Eqs.~\eqref{eq:rh-hayward}  or \eqref{eq:rh-minkowski}.  We will find that the new $M$-dependence that we have introduced  leads to substantially novel thermodynamic behavior.

\begin{figure}[!htb]
\centering
\includegraphics[width=0.48\textwidth]{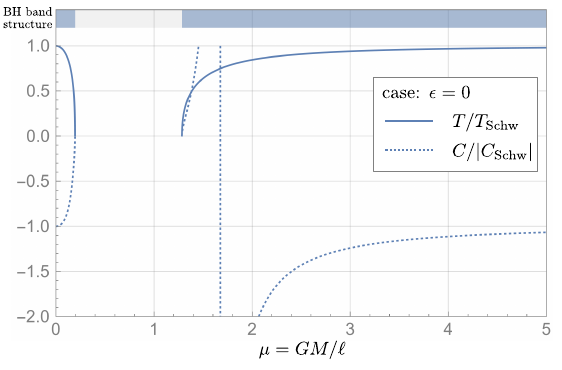}
\includegraphics[width=0.48\textwidth]{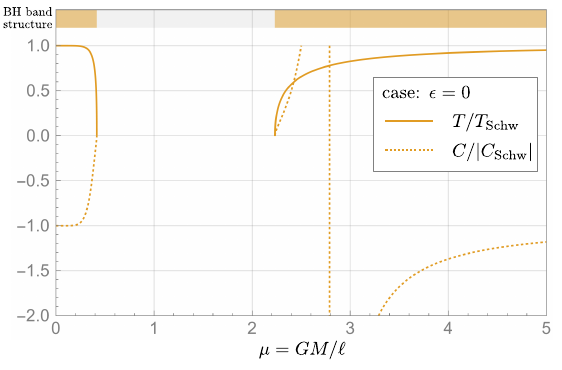}
\caption{For $\epsilon=0$, we visualize the Hawking temperature normalized to the Schwarzschild value (solid line) as well as the specific heat normalized to the absolute Schwarzschild value (dashed line) as a function of the dimensionless mass parameter $\mu = GM/\ell$. The left panel shows the de\,Sitter core case, and the right panel shows the Minkowski core case. The black hole band structure is highlighted as a solid bar in the upper part of each panel. For masses outside this region, displayed in light gray, no black holes exist. This figure is shown as a point of comparison as the limiting curvature condition requires $\epsilon$ to be nonzero.}
\label{fig:thermo-epsilon=0}
\end{figure}

\begin{figure}[!htb]
\centering
\includegraphics[width=0.96\textwidth]{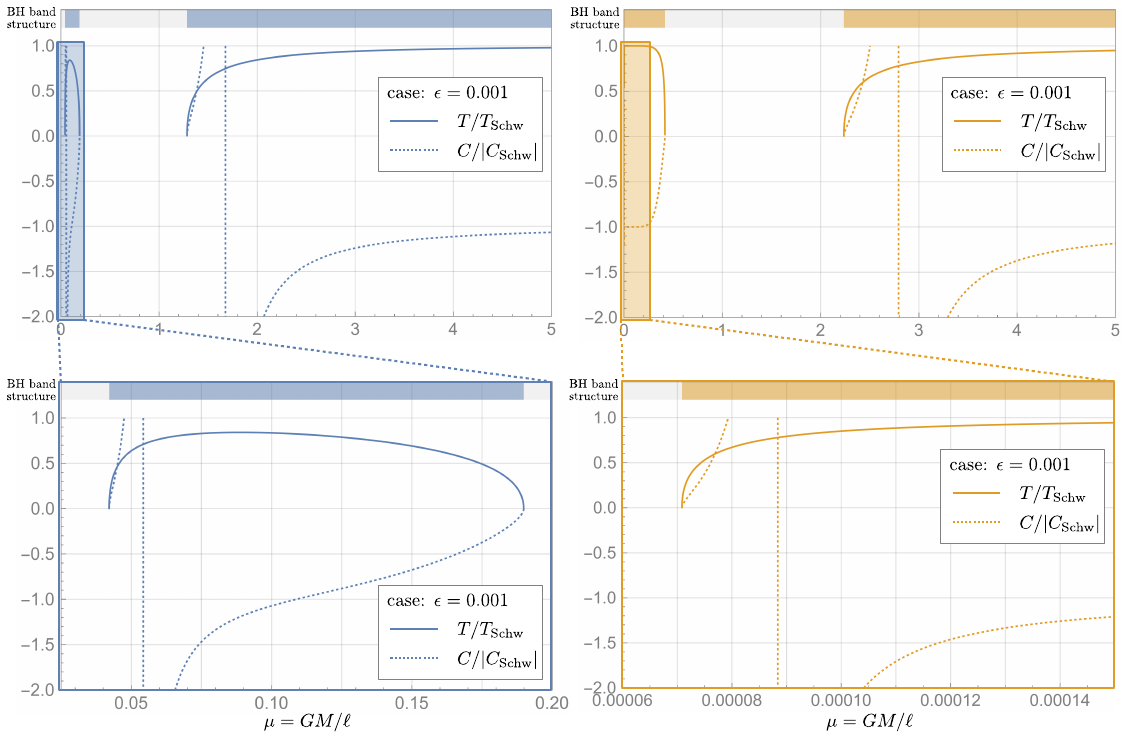}
\caption{For $\epsilon=0.001$, we visualize the Hawking temperature and specific heat normalized for the de\,Sitter core case (left) and the Minkowski core case (right) in an identical fashion to Fig.~\ref{fig:thermo-epsilon=0}. The non-trivial band structure induced by $\epsilon > 0$ is visualized via a zoomed-in view in the second row, highlighting the existence of a mass gap in the lower black hole band. In both cases, the qualitative behavior of the Hawking temperature and specific heat is identical to the behavior at the lower end of the second part of the black hole band spectrum.}
\label{fig:thermo-epsilon=0.001}
\end{figure}

Since our explicit expressions for $T$ and $C$ are rather cumbersome, we only display our results graphically.  To do so, we
keep the regulator $\ell$ fixed and consider its effect on a range of black hole masses $M$. Hence it is convenient to define the dimensionless quantity
\begin{align}
\mu \equiv \frac{GM}{\ell} = \frac{1}{2\, \hat{\ell}} \, .
\end{align}
We may then plot $T_H$ normalized to the Schwarzschild Hawking temperature as well as the heat capacity normalized to the absolute value of the Schwarzschild heat capacity, both as a function of $\mu$. We plot these thermodynamical quantities for the case of $\epsilon=0$ (Fig.~\ref{fig:thermo-epsilon=0}) as well as for $\epsilon=0.001$ (Fig.~\ref{fig:thermo-epsilon=0.001}), while highlighting the black hole band structure as we vary $\mu$.  We note that Fig.~\ref{fig:thermo-epsilon=0} is presented only as a point of comparison, since the choice $\epsilon=0$ does not satisfy the limiting curvature condition.
 
Notably, in the case where $\epsilon$ is nonzero, we find that the edges of the black hole bands generically feature vanishing Hawking temperatures with vanishing heat capacity. Moreover, the heat capacity approaches zero from above at the lower end of a given black hole band, and from below at the upper end of a black hole band. This implies that the black holes at the low-mass end of a black hole band can be interpreted as thermodynamically stable end products of Hawking evaporation from larger black holes from within the same band.\footnote{The potential role of such remnants as a part of dark matter has been addressed in e.g.~Refs.~\cite{Dymnikova:2015yma,Trivedi:2025vry}.} This is to be starkly contrasted with the case of Schwarzschild black holes, which always decay, and which exist for all masses. Similarly,  the more conventional  de\,Sitter core or Minkowski core black holes discussed in the literature do not feature low-mass bands. Significantly, for $f \not=1$ and $\epsilon\not=0$ the mass range of thermodynamically stable and potentially viable dark matter mass windows is appreciably enlarged.
 
\begin{figure}[!htb]
\centering
\includegraphics[width=0.48\textwidth]{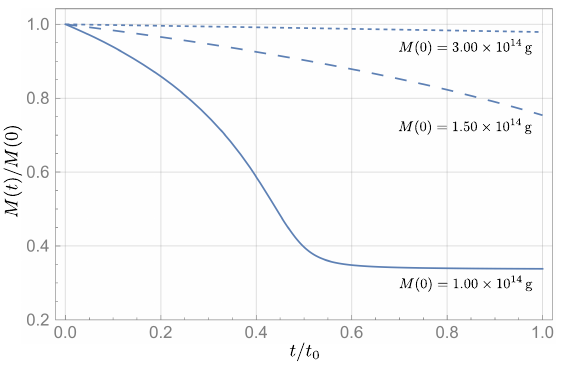}
\includegraphics[width=0.48\textwidth]{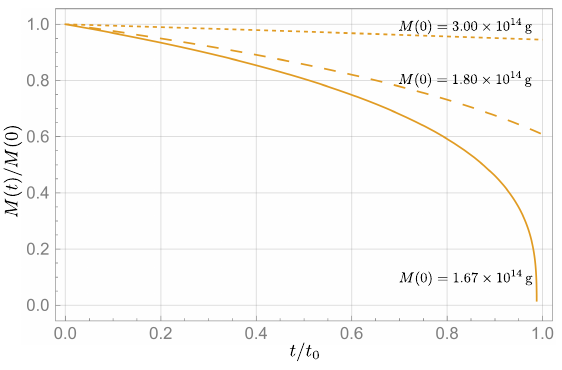}
\caption{Given a choice of initial regulator $\hat{\ell}_0= \ell/(2GM_0)$ and black hole model, we plot the evolution of the normalized black hole mass as a function of dimensionless time (measured in the age of the Universe $t_0$). The left panel shows the case of de\,Sitter core black holes with $\hat{\ell}_0 = 4$ and $\epsilon=0.001$, and the right panel depicts the case of the Minkowski core black holes with $\hat{\ell}_0 = 2$ and $\epsilon=0.001$. For the masses greater than approximately $10^{15}$~g, which are considered throughout the rest of this paper, the black hole mass loss is hence small enough that their masses may be considered constant over the lifetime of the Universe.}
\label{fig:lifetime}
\end{figure}

 To obtain a simple estimate for the black hole lifetime, we model the evaporation via the Stefan-Boltzmann law, ignoring greybody factors:
\begin{equation}
\label{eq:evaporation}
\frac{\dd M}{\dd t} = -\frac{1}{c^2} \, \sigma \, 4\pi r_H^2 \, T_H^4 \, , \quad
\sigma = \frac{2\pi^5 k_B^4}{15 c^2 (2\pi \hbar)^3} \, ,
\end{equation}
Here $\sigma$ is the Stefan-Boltzmann constant and $4\pi r_H^2$ is the spatial area of the black hole event horizon.  Working in dimensionless mass $\mu = GM/(c^2 \ell)$, time $\tau = ct/\ell$, radius $\rho_H = r_H / \ell$,  and temperature $\Theta_H = k_B \ell/(\hbar c) T_H$ the above takes the suggestive form 
\begin{align}
\frac{\dd\mu}{\dd\tau} = - \frac{\pi^3}{60} \left(\frac{m_\text{Pl}}{M_0}\right)^2 \frac{1}{\hat{\ell}^2_0} \, \rho_H^2 \Theta_H^4 \, ,
\label{eq:mutau}
\end{align}
where $M_0$ is the initial black hole mass and  $\hat{\ell}_0 \equiv c^2 \ell / (2 G M_0)$.  Equation~(\ref{eq:mutau}) can easily be integrated numerically. We thereby extract the mass evolution curves depending on the initial mass $M_0$ of the black hole; see Fig. \ref{fig:lifetime}. In this figure we keep $\hat{\ell}_0$ fixed, implying that the depicted curves correspond to slightly different magnitudes for the regulator $\ell$. While this choice has little effect in this figure, it is made for consistency with our analysis in Sec.~\ref{sec:gamma}: there we restrict our attention to a slice of parameter space with constant $\hat{\ell}$, where black holes over a wider range of masses are expected to have regulator effects at the horizon that are of comparable importance.

Our estimate suggests that primordial black holes will experience continuous energy loss through Hawking radiation, approaching an asymptotic lower bound which represents the thermodynamically stable end product of the respective mass band.   However, our results indicate that  the black hole mass loss is small enough when 
$M \agt 10^{15}$~g, that an approximation of constant mass during the lifetime of the universe is justified. The masses that we consider in the remainder of this paper will fall in the range $10^{15}$ to $10^{17}$~g.

\section{Dark Matter Fraction} \label{sec:gamma}

The Hawking radiation from primordial black hole dark matter was taken into account in our discussion of the black hole lifetimes in the previous section.  This radiation has potentially observable consequences;  the experimental measurements of the extragalactic gamma ray flux over a range of energies allows one to place a bound on the fraction of the total dark matter that can be made up of primordial black holes.  The relevant analysis is relatively standard and has been used in the study of other possible primordial black hole dark matter candidates~\cite{Calza:2024fzo,Calza:2024xdh,Calza:2025mwn}.   We consider this issue in terms of the primordial black holes of interest to us, primarily as an existence proof that dark matter of the type of interest can make up a substantial fraction of the total amount of dark matter.

The number of particles of species $i$ and spin $s$ emitted by a black hole via Hawking radiation per unit energy and per unit time is given by
\begin{equation}
\frac{\dd^2 N_i}{\dd t \, \dd E_i} = \frac{n_i}{2 \pi} \sum_{l,m} \frac{\Gamma^s_{l,m}(\omega)}{e^{E_i/T} \pm 1} \,\, ,
\label{eq:spec}
\end{equation}
working in units where $\hbar$, $c$ and the Boltzmann constant $k_B$ have been set to $1$.   The quantity $n_i$ is the number of degrees of freedom of the particle in question (for example, $n_i=2$ for a massless gauge field with two physical polarizations) and
the sign in the denominator is $-$ ($+$) for bosons (fermions).   The greybody factors $\Gamma^s_{l,m}(\omega)$, are a function of the particle energy (i.e., $\omega = E_i / \hbar$ in conventional units) which captures how the spectrum measured asymptotically far from the black hole horizon differs from that of a blackbody; $l$ and $m$ are the angular momentum labels for the partial wave decomposition of this radiation.   The greybody factors may be related to the solutions of the radial Teukolsky equation~\cite{Teukolsky:1973ha}, which determines how the massless perturbations of different spins propagate in the black hole background.   We will focus on the photons that are emitted via Hawking radiation.  While the computation of the greybody factors is clearly presented in Ref.~\cite{Calza:2024fzo}, there are a few typographical errors in some of the equivalent forms of the radial Teukolsky equation that are displayed in that reference;  to avoid any confusion, we present the formulae necessary for computing the greybody factors in appendix \ref{app:greybody} using our notation. 

Denoting quantities that are measured in the present epoch by the subscript $0$, the present number density of photons with energy $E_{\gamma 0}$ may be written
\begin{equation}
n_{\gamma 0}(E_{\gamma 0}) = n_\text{pbh}(t_0) \, E_{\gamma 0} \int_0^{z_\star} \frac{\dd z}{H(z)}\, \left. \frac{\dd^2 N_\gamma}{\dd t \, \dd E_\gamma}\right|_{E_\gamma=(1+z) E_{\gamma 0}} \,\, ,
\end{equation}
where $n_\text{pbh}(t_0)$ is the present number density of primordial black holes with mass $M$, and the integral sums over photons emitted at different times, taking into account their cosmological redshift.  The limit $z_\star$ represents the redshift at recombination, and $H(z)$ is the Hubble parameter.  Ignoring the radiation domination era we parametrize
\begin{align}
H(z) = H_0 \, \sqrt{\Omega_\Lambda + \Omega_\text{m} (1+z)^3 } \, ,
\end{align}
where used the numerical values $H_0 = 67.37 \frac{\text{km}}{\text{s}\,\text{Mpc}}$, $\Omega_\Lambda = 0.6847$, $\Omega_\text{m} = 0.3146$ (with $\Omega_\text{dm} = 0.2640$), and $z_\star = 1089.92$ extracted from the Planck 2018 results \cite{Planck:2018vyg}.

The examples we consider assume a monochromatic spectrum of black hole masses, and we focus on $M \agt 10^{15}$~g where we have shown that the change in mass due to Hawking radiation is negligible over the age of the Universe.  (A more general treatment will be considered elsewhere.) The number of photons per unit time per unit area per unit solid angle is then
\begin{equation}
I(E_{\gamma 0}) = \frac{c}{4 \pi} \, n_{\gamma 0}(E_{\gamma 0}) \,\, ,
\end{equation}
which can be compared with observations to determine an upper bound on $n_\text{pbh}(t_0)$.  Once that bound is determined, the fraction of the total amount of dark matter that is primordial black holes may also be bounded
\begin{equation}
f_\text{pbh} = \frac{\Omega_\text{pbh}}{\Omega_\text{dm}} = \frac{n_\text{pbh}(t_0) \, M }{\rho_{c0} \, \Omega_\text{dm}} \,\, ,
\end{equation}
where $\rho_{c0} \equiv 3 H_0^2 / (8 \pi G)$ is the current critical density.

\begin{figure*}[!htb]
\centering
\includegraphics[width=0.75\textwidth]{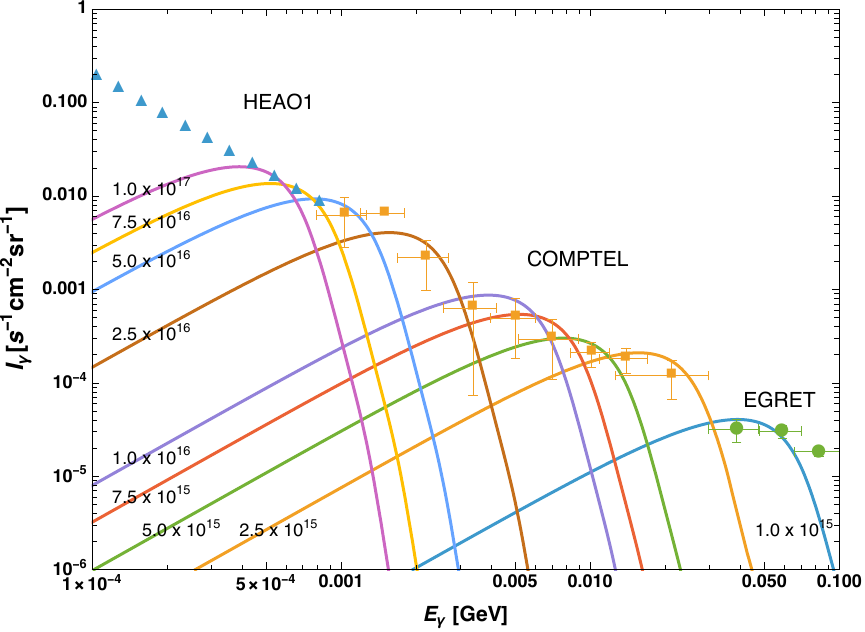}
\caption{Primary photon flux (solid lines) versus the extragalactic gamma ray background (points).  The observational data are discussed in the text.  The numbers labeling each curve are the primordial black hole masses in grams, assuming the de\,Sitter core case with $\epsilon=0.001$.
}
\label{fig:flux}
\end{figure*}
\begin{figure*}[!htb]
\centering
\includegraphics[width=0.47\textwidth]{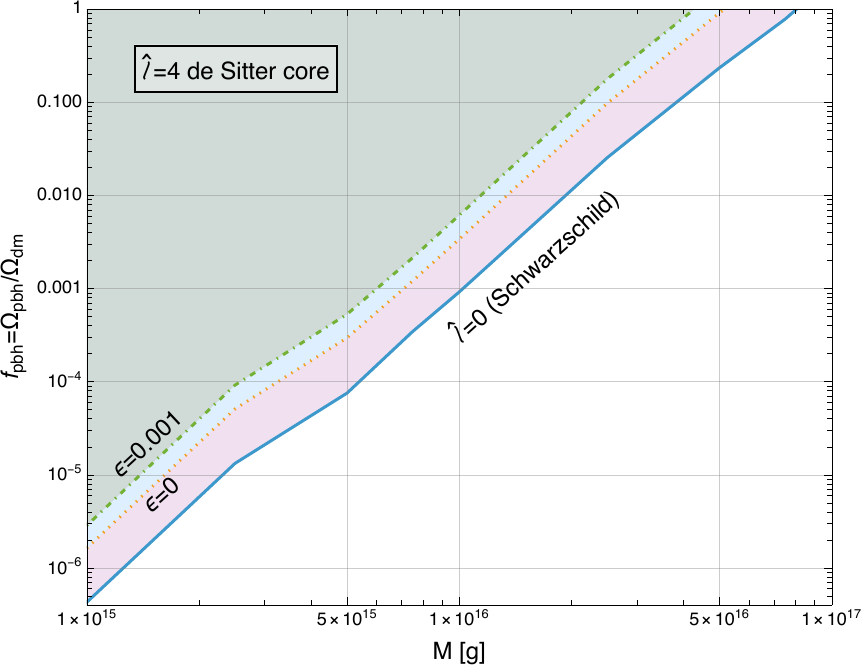} \hspace{1em}
\includegraphics[width=0.47\textwidth]{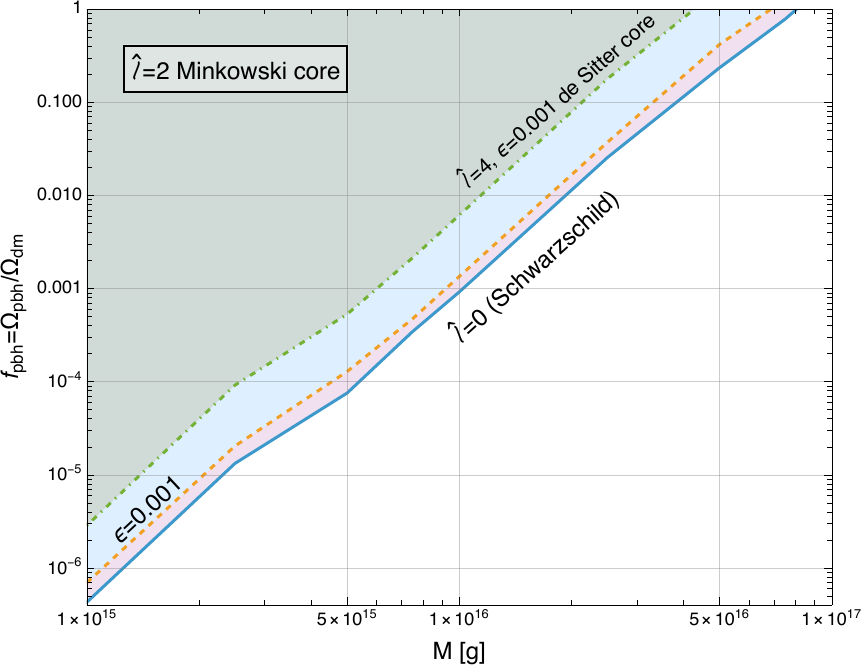}
\caption{Excluded values (shaded) of the fraction of dark matter that is made up of primordial black holes.  The left panel considers the de\,Sitter core case with 
$\hat{\ell}=4$; the dot-dashed and dotted lines correspond to the choices $\epsilon=0.001$ and $0$, respectively.  The right panel shows the excluded regions for the Minkowski core case with $\hat{\ell}=2$ (dashed line).  The dot-dashed line again represents the de\,Sitter core $\hat{\ell}=4$, $\epsilon=0.001$ line from the left panel, provided for easier comparison.}
\label{fig:exclude}
\end{figure*}

As an example, we show in Fig.~\ref{fig:flux} the photon flux for a range of de\,Sitter core black hole masses, corresponding to the choice of $f(\hat{\ell})$ in Eq.~(\ref{eq:fdef}) with $\epsilon=0.001$.  The curve corresponding to each choice of $M$ has $n_\text{pbh}(t_0)$ adjusted so that the flux does not exceed 
the $1\,\sigma$ error bars on any data point for the observed flux from three observatories (the same considered in Ref.~\cite{Calza:2024fzo}), namely HEAO-1~\cite{Gruber:1999yr}, COMPTEL~\cite{Schoenfelder:2000bu} and EGRET~\cite{Strong:2004ry}.   The analogous plot for the Minkowski core black holes (with $\epsilon=0$) is qualitatively similar, but is not shown.  The resulting upper bounds on $n_\text{pbh}(t_0)$ may be used to determine the excluded regions of the $f_\text{pbh}$ versus $M$ plane.   Characteristic examples are shown in Fig.~\ref{fig:exclude}:   The left panel compares the excluded fraction for the modified de\,Sitter core black holes, previously defined, to the $\ell=0$ Schwarzschild limit.   The plot assumes $\hat{\ell}=4$, with the difference between the dot-dashed and dotted lines indicating how the bound changes as the parameter $\epsilon$ is varied from $0.001$ (dot-dashed line) to zero (dotted line).   The right panel compares the excluded fraction for modified Minkowski core black holes, previously defined, to the $\ell=0$ Schwarzschild limit.  The bound for the Minkowski core black hole is indicated by the dashed line, assuming $\hat{\ell}=2$;  the dot-dashed line is again the de\,Sitter core black hole with $\hat{\ell}=4$ and $\epsilon=0.001$, which is provided for comparison.  These plots show that a larger fraction of primordial nonsingular black holes are allowed at a given mass compared to the Schwarzschild case, similar to the other nonsingular black holes that were studied in Ref.~\cite{Calza:2024fzo}. The effect of the regulator
is less pronounced in the Minkowski core case due to the faster fall off of $(2 G M \ell^2)^{1/3} f(\hat{\ell})$ at large $\ell$ compared to $\ell^2 f(\hat{\ell})$ in the de\,Sitter core case.

\section{Conclusions} \label{sec:conc}

Metrics that describe black holes in general relativity have physical singularities that are usually assumed to indicate a breakdown in the classical theory.  Phenomenological metrics that involve simple modifications to remove the singularities are hoped to capture some of the effects of the ultraviolet physics that may remedy the problem, allowing observational consequences to be considered before a full theory is known.   Here we have studied a modification to some popular nonsingular black hole metrics that introduces an additional mass dependence which allows the regulator scale to be comparable to or exceed the horizon radius, and leads to new mass ranges where black holes are obtained rather than compact, horizonless objects.  In the case where these black holes have a de\,Sitter core, our metrics have sub-Planckian curvatures both at the origin and where the curvature invariants assume their maximum values.  In the case where the black holes have Minkowski cores, the same is true, with the curvature vanishing at the origin.  In the models we propose, the limiting curvature condition is satisfied for all values of the black hole mass.   

While our previous work considered kilometer-scale regulators and potentially observable effects in stellar-mass black holes, we have looked here at femtometer scale regulators in the nonsingular black holes of interest,  which can nonetheless be larger than the horizon scale for primordial black holes that may serve as candidates for dark matter.   Significant horizon-scale effects include the modification of thermodynamic quantities that depend on derivatives of $M$, so that the mass dependence we introduce cannot be interpreted as a simple rescaling of the regulator parameter.  One of the consequences is the emergence of novel thermodynamically stable black hole branches discussed in Sec.~\ref{sec:thermo}. In addition to considering the new black hole mass ranges, thermodynamics and lifetimes, we computed the primary photon flux due to Hawking radiation from these black holes.  Using data on the observed extragalactic gamma ray background, we determined upper bounds on the fraction of the dark matter density that can be made up of such primordial nonsingular black holes, providing support for the viability of the scenario.

The present work establishes, for some choices of the black hole masses and the regulator function $f(\hat{\ell})$, that the nonsingular black hole metrics we discuss are among the set of possible candidates that might describe primordial black hole dark matter.   The present analysis could be generalized by considering other forms for the regulator function $f(\hat{\ell})$,  other mass ranges in which the black hole mass would be expected to change appreciably through Hawking radiation over the lifetime of the universe, and other assumptions about the initial black hole mass spectrum.   It would also be of interest to see whether a plausible effective description of quantum gravity can be found that leads to metrics of the form we discuss as exact solutions and that might also suggest a preferred form for the regulator function.  These topics will be considered in future work.
\begin{acknowledgments} 
SA and CDC thank the NSF for support under Grant No. PHY-2112460 and No. PHY-2411549; JB is grateful for support as a Fellow of the Young Investigator Group Preparation Program, funded jointly via the University of Excellence strategic fund at the Karlsruhe Institute of Technology (administered by the federal government of Germany) and the Ministry of Science, Research and Arts of Baden-W\"urttemberg (Germany).
\end{acknowledgments}

\appendix
\section{Greybody factors}
\label{app:greybody}
Following Ref.~\cite{Calza:2024fzo}, the greybody factors may be related to the coefficient of the incoming wave solution to the radial Teukolsky equation~\cite{Teukolsky:1973ha}, describing a perturbation of spin $s$.   Working with the variable 
\begin{equation}
x \equiv (r-r_H)/r_H \,\, ,
\end{equation}
where $r_H$ is the horizon radius, the radial Teukolsky equation can be written in the form
\begin{equation}
A(x) \, \ddot{R}_s + B(x) \, \dot{R_s} +C(x) \, R_s=0 \,\, ,
\label{eq:teul}
\end{equation}
where dots represent derivatives with respect to $x$, and the coefficient functions are
\begin{align}
A(x) &= (x+1)^2 F^2 \,\, ,\label{eq:A} \\
B(x) & = (s+1) \, F^2 \left[ 2\,(x+1)+(x+1)^2 \dot{F}/F \right] \,\, , \label{eq:B}\\
C(x) &= (x+1)^2 \omega^2\, r_H^2 + 2 \,i \,s \, \omega \,r_H (x+1) \, F - i \,s \,  \omega \, r_H \, \dot{F} (x+1)^2 \nonumber \\
&+s F \left[(x+1)^2 \ddot{F}+4(x+1)\dot{F}+2F-2\right] +\left[s(s+1)-l(l+1)\right] F \,\, . \label{eq:C}
\end{align}
We note that the function $B(x)$ disagrees with the one given in Ref.~\cite{Calza:2024fzo}; nevertheless, our numerical
results agree with what is presented in that reference, so we believe the error is typographical.   For the de\,Sitter core black holes, the metric function
$F(x)$ is
\begin{equation}
F(x) = 1 - \frac{r_H^2 \, (x+1)^2}{\ell^2 f(\hat{\ell})+[r_H^2-\ell^2 f(\hat{\ell}) ](x+1)^3 }   \,\,\, ,
\end{equation}
and for the Minkowski core black holes defined by Eq.~(\ref{eq:MCF}) one has 
\begin{equation}
F(x) = 1 -\frac{1}{x+1} \exp\left[\frac{x}{x+1}  \frac{\hat{\ell}^{2/3} f(\hat{\ell})}{\hat{r}_H}\right] \,\,\, .
\end{equation} 
One can show by substitution that Eqs.~(\ref{eq:teul})--(\ref{eq:C}) have a solution near the horizon ($x \ll 1$) of the form
\begin{equation}
R_s(x) = x^{-s-i \, \frac{\omega\, r_H}{ \tau}} \sum_{n=0}^\infty a_n \, x^n \,\, ,
\end{equation}
where $\tau$ is given by
\begin{equation}
\tau = \left\{ \begin{array}{ll} \displaystyle 1-3\frac{\hat{\ell}^2 f(\hat{\ell})}{\hat{r}_H^2} & \hspace{2em} \mbox{de\,Sitter core,} \\[10pt]
\displaystyle 1-\frac{\hat{\ell}^{2/3} f(\hat{\ell})}{\hat{r}_H} & \hspace{2em} \mbox{Minkowski core.} \end{array} \right.
\end{equation} 
The greybody factor is found by using the near horizon solution normalized such that $a_0=1$ to integrate Eq.~(\ref{eq:teul}) out to
large radius where it can be compared to the asymptotic form
\begin{equation}
R_s(x) \rightarrow R_s^{in}(\omega) \, \frac{e^{-i \, \omega\,r_H x}}{x} + R_s^{out}(\omega) \, \frac{e^{i \, \omega\,r_H x}}{x^{2s+1}}. \,\,\ .
\end{equation}
The greybody factor is then given by~\cite{Calza:2024fzo}
\begin{equation}
\Gamma^s_{lm}(\omega) =  \frac{i\, \tau\,e^{i\pi s} \, (2 \,\omega\, r_H)^{2 s-1} \Gamma(1-s-\frac{2 \, i\,  \omega\,  r_H}{\tau})}{\Gamma(s-\frac{2 \, i \, \omega \, r_H}{\tau})}
\,  | R_s^{in}(\omega)|^{-2} \,\, .
\end{equation}
This is used in the numerical evaluation of Eq.~(\ref{eq:spec}).

\pagebreak

\end{document}